\documentclass[12pt]{iopart}
\usepackage{iopams}
\usepackage{graphicx}

\newcommand{\bsy}[1]{\boldsymbol{#1}}

\begin{document}
\title[Graphene Rings in Magnetic Fields]{Graphene Rings in Magnetic Fields: \\ Aharonov-Bohm Effect and Valley Splitting}

\author{J Wurm$^{1,2}$, M Wimmer$^{1}$, H U Baranger$^{2}$ and K Richter$^{1}$}
\address{$^1$Institut f\"ur Theoretische Physik, Universit\"at Regensburg, D-93040, Germany}
\address{$^2$Department of Physics, Duke University, Box 90305,
Durham, North Carolina 27708-0305, U.S.A.}
\ead{Juergen.Wurm@physik.uni-regensburg.de}

\begin{abstract}
  We study the conductance of mesoscopic graphene rings in the
  presence of a perpendicular magnetic field by means of numerical
  calculations based on a tight-binding model. First, we consider the
  magnetoconductance of 
 such rings and observe the Aharonov-Bohm effect. 
  We investigate different regimes of the magnetic flux up to
  the quantum Hall regime, where the Aharonov-Bohm oscillations are
  suppressed. Results for both clean (ballistic) and disordered (diffusive) 
  rings are presented.
  Second, we study rings with smooth mass
  boundary that are weakly coupled to leads. We show that the valley
  degeneracy of the eigenstates in closed graphene rings can be lifted
  by a small magnetic flux, and that this lifting can be
  observed in the transport properties of the system.
\end{abstract}

\section{Introduction}

Since their experimental discovery, graphite monolayers, also known as
graphene, have attracted a huge amount of interest among both
experimentalists and theorists due to the linear low energy dispersion
and various properties stemming from this unusual dispersion \cite{Geim2007,
CastroNeto2009}. For instance, graphene has opened new perspectives for mesoscopic
physics, such as pseudodiffusive conductance at the Dirac point
\cite{Tworzydlo2006,Danneau2008,DiCarlo2008}, specular
Andreev reflection \cite{Beenakker2008}, or the signatures 
of symmetries of the graphene Hamiltonian in the conductance 
of diffusive wires \cite{Suzuura2002, Khveshchenko2006, McCann2006, Morpurgo2006, Aleiner2006} and ballistic cavities
\cite{Ponomarenko2008, Wurm2009}.

Mesoscopic rings can be considered as prototypical devices in 
mesoscopic physics, as they show one of the most basic coherence effects,
namely the Aharonov-Bohm (AB) effect \cite{Aharonov1959, Webb1985, Washburn1986}:
oscillations of the transmission, or dimensionless conductance, $T= (h /2e^2) G$ as a
function of the magnetic flux $\Phi$ through the ring.  The reason for
these oscillations is the phase difference $\Delta\varphi = 2\pi \Phi
/ \Phi_0$ between electrons travelling along the different arms of the
ring. Here, $\Phi_0 = h/e$ is the magnetic flux quantum. Fifty years after its discovery, 
the AB effect is one of the most well-known manifestations of quantum interference
within and beyond the field of condensed matter physics.

Hence it is rather surprising that the AB effect in graphene has up to now
received only a little attention in the literature. Most notably, there
are only two experiments on graphene AB rings manufactured by electron beam 
lithography \cite{Russo2008,Molitor2009}, one of them  leaving many open questions on the physical
origin of some of the observed effects \cite{Russo2008}. From the theory side, there is
only one numerical study of the AB effect in graphene rings; it focuses 
on the effects of valley-polarized currents, i.e.~on the
few-mode or low-doping regime in the leads \cite{Rycerz2009}. 
In this work, we will in contrast also consider the many-mode or
high-doping regime.

In addition to these studies focusing on the transport
properties of \emph{open} rings, there has been a proposal
to use the Aharonov-Bohm effect in \emph{closed} rings to form qubits:
The energy spectrum of a closed graphene
ring with infinite mass boundary condition \cite{Berry1987} has been
calculated in reference \cite{Recher2007}, where the authors find that
the valley degeneracy of the energy levels is lifted as soon as a
magnetic flux pierces the ring. This effect has also been found
for chaotic graphene quantum dots \cite{Wurm2009}. Note that this aspect 
is not present
in AB rings realized e.g.~in semiconductor heterostructures and
metals. It is connected to
a special antiunitary symmetry of the Dirac Hamiltonian,
which describes graphene well for low Fermi energies. In this work, we
will show that the lifting of the valley degeneracy is also
visible in the transport properties of graphene rings.

The paper is organized as follows:
In the first part, we investigate the AB effect of
graphene structures by numerically calculating the transmission of
rings attached to infinitely extended leads. We study both small rings
in the one-mode regime and large rings with many modes propagating in both
the leads and arms of the ring. In the latter we especially consider the 
high-field regime and the effects of disorder.  In the
second part of this work, we show that the breaking of valley-degeneracy 
by a magnetic field is also visible in the transport properties
of graphene rings. We do this by numerically calculating the transmission of graphene rings
that are weakly coupled to two leads. This transmission shows peaks as a function
of the Fermi energy $E_F$ which correspond to the energy levels
of a closed ring; the lifting of their degeneracy can be observed
as a splitting of the transmission peaks upon applying a magnetic field
perpendicular to the ring.

For our numerical work, we use a tight binding model taking into
account the $2p_z$-orbitals of the carbon atoms, leading to the
Hamiltonian
\begin{equation}
H_{\mathrm{tb}} = \sum_{\langle i,j \rangle} t_{ij}\, c_i^{\dagger}c_j + \sum_{i} m_i\, c_i^{\dagger}c_i
\end{equation}
with $i$ and $j$ beeing nearest neighbor sites in the first sum. The
magnetic field is included via the Peierls substitution $t_{ij} =
-t\exp\left(i\frac{2\pi}{\Phi_0}\int_{\bsy{r}_i}^{\bsy{r}_j}\bsy{A}\cdot
\bsy{dr}\right)$. The second term accounts for a staggered on-site
potential, i.\,e.\ $m_i = m(\bsy{r_i})$ is positive (negative) if
$\bsy{r_i}$ is located on sublattice A (B). Such a staggered potential
corresponds to a mass term in the effective Dirac equation and will be
used in the second part of this paper to suppress the inter-valley
scattering that breaks the valley degeneracy \cite{Wurm2009}. The
lattice points are determined by cutting a ring out of the graphene
sheet [cf. figure \ref{fig:ringScheme2}\,(b)]. In order to solve the transport problem to obtain the
dimensionless conductance $T$ within this tight-binding model, we use
an adaptive recursive Green function method \cite{Wimmer2008a}.

\section{Aharonov-Bohm effect in graphene rings}
\begin{figure}
\centering
\includegraphics[clip, width=9.0cm]{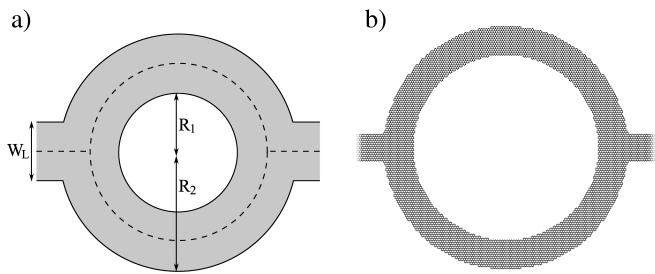}
\caption{(a) Schematic of the systems studied numerically. The parameters
  defining the shape are the inner and outer radius $R_1$ and $R_2$,
  respectively, and the width $W_L$ of the infinitely extended leads.
  The dashed line marks the points where the mass term used in Section
  \ref{sec:Vdeg} is zero. (b) Example of a system defined by cutting a ring out of a graphene sheet 
[the system was used for figure \ref{fig:SmallAB} (a) and (c)]. }
\label{fig:ringScheme2}
\end{figure}

We now investigate the transmission of graphene rings with
two attached leads under the influence of a perpendicular magnetic
field $B$, giving rise to a flux $\Phi$ through the ring. In the
following we define $\Phi = B\,\pi \bar{R}^2$ as the flux through a
circle with the average ring radius $\bar{R}=(R_1+R_2)/2$ (cf.
figure \ref{fig:ringScheme2}). Because electrons can travel along one or
the other arm of the ring, the transmission $T(\Phi)$ is expected to
oscillate with periods $\Delta \Phi = \Phi_0/n, n\in \mathbb{N}$, as
mentioned in the introduction. The reason why more than one single period, namely
$\Phi_0$, may occur is that the electrons do not necessarily have to
leave the ring after traversing an arm once, but rather they may circle around the ring several times before exiting, giving rise to
higher harmonics with $n>1$. 

\begin{figure}
\centering
\includegraphics[clip, width=9.0cm]{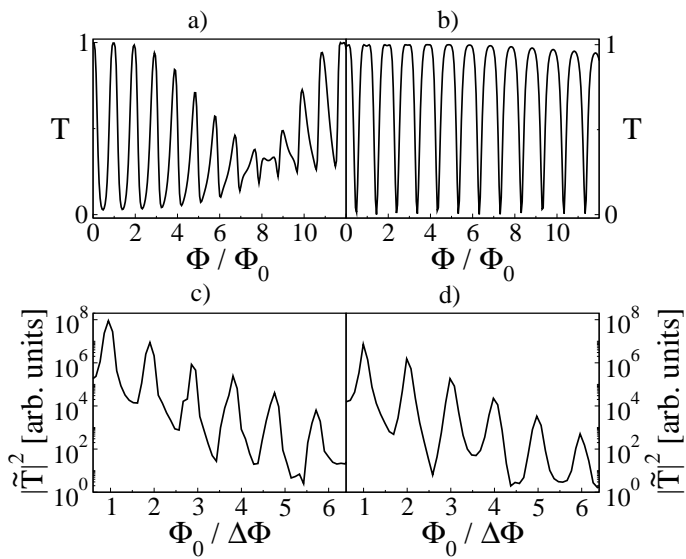}
\caption{Aharonov-Bohm oscillations [(a),(b)] and corresponding power spectra [(c),(d)] of small rings in the one-mode regime. (a),(c) Ring with zigzag type leads and
$R_1=9.84$\,nm, $R_2=13.0$\,nm, $W_L=3.20$\,nm. (b), (d) Ring with armchair type leads and $R_1=9.84$\,nm, $R_2=12.5$\,nm, $W_L=4.92$\,nm. $\tilde{T}(1/\Delta\Phi)$ is the Fourier transform of $T(\Phi)$ where $\Delta\Phi$ is the period of the oscillation related to the Fourier transform variable by $k=2\pi/\Delta\Phi$.}
\label{fig:SmallAB}
\end{figure}

We begin by considering small graphene rings in the one-mode regime.
In figure \ref{fig:SmallAB}\,(a) and (b) we show the numerically
calculated transmission as a function of $\Phi$ for two small rings
with radii of about $10$\,nm and widths of about $3$\,nm (see figure
caption for details). One of the rings has zigzag type leads while the
other has armchair type leads. For both, one can clearly see the AB
oscillations with period $\Phi=\Phi_0$.  To expose a few more details
of the frequency content of our data, we show the power spectra of
the AB oscillations in figure \ref{fig:SmallAB}\,(c) and (d).  There
we find pronounced peaks for the fundamental frequency and for several
higher harmonics (we show only the first five). Note that we plot
the spectra on a logarithmic scale, since the fundamental frequency
is strongly dominating over the higher ones. The peaks in the spectrum
lie very close to multiple values of $1/\Phi_0$. Because of the finite
width of the rings, one has certain allowed deviations from these
values. For example, for the system of figure \ref{fig:SmallAB}\,(a), the
fundamental frequency is expected to lie between $0.74/\Phi_0$ and
$1.3/\Phi_0$, obtained from the inner and outer ring radii $R_1$
and $R_2$, respectively. The fact that the peaks are slightly shifted to
frequencies lower than $n/\Phi_0$ means that the effective radius of
our rings is slightly smaller than the average radius $\bar{R}$.
We also performed calculations for various systems with different parameters and 
found the same behavior irrespective of the details like radii, width or lead type.

Next, we turn to the case of larger graphene rings in the many-mode regime, i.e.~for
high values of doping. This is the regime applicable to the available experiments
of Refs.~\cite{Russo2008,Molitor2009}. In the numerical simulations we consider
rings with an average radius $\bar{R}\approx 55\,$nm. While this size is
still smaller than in the experiments ($\bar{R}\approx 500\,$nm in Ref.~\cite{Russo2008}
and $\bar{R}\approx 250$-$300\,$nm in Ref.~\cite{Molitor2009}), 
the rings contain more than a hunded thousand atoms
and hence can be considered as mesoscopic objects. Therefore,
we do not expect that larger rings would show any fundamentally different physical properties
than the systems considered here.

\begin{figure}
\centering
\includegraphics[clip, width=11.0cm]{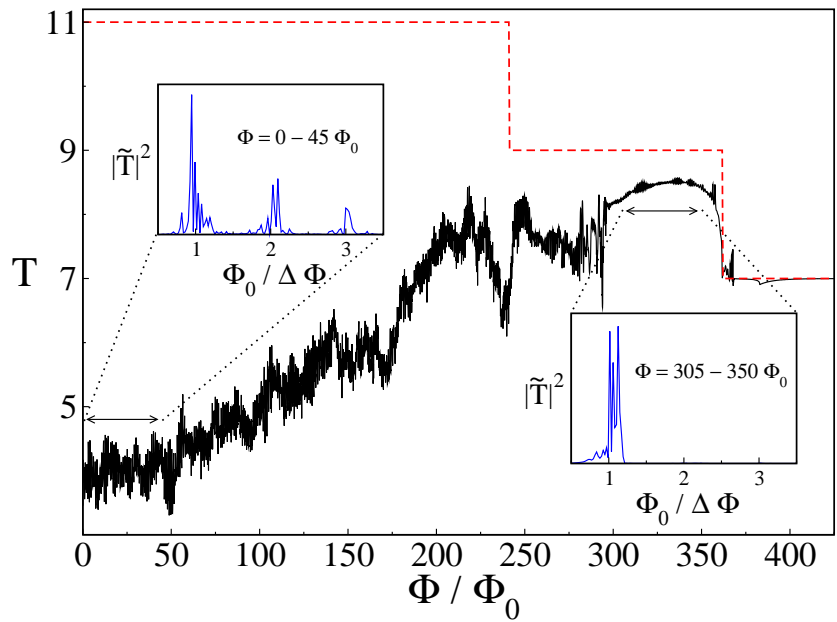}
\caption{Magnetotransmission of a large ring with $R_1=45.5$\,nm, $R_2=63.7$\,nm and $W_L=13.5$\,nm for a wide range of the magnetic flux $\Phi$ (solid black line) 
in the multi-mode regime. The
red dashed line shows the number of occupied transverse channels in the leads. The insets show power spectra for different regimes: left inset $\Phi= 0$-$45\,\Phi_0$, right inset $\Phi= 305$-$350\,\Phi_0$.}
\label{fig:ABlarge185}
\end{figure}

\begin{figure}
\centering
\includegraphics[clip, width=11.0cm]{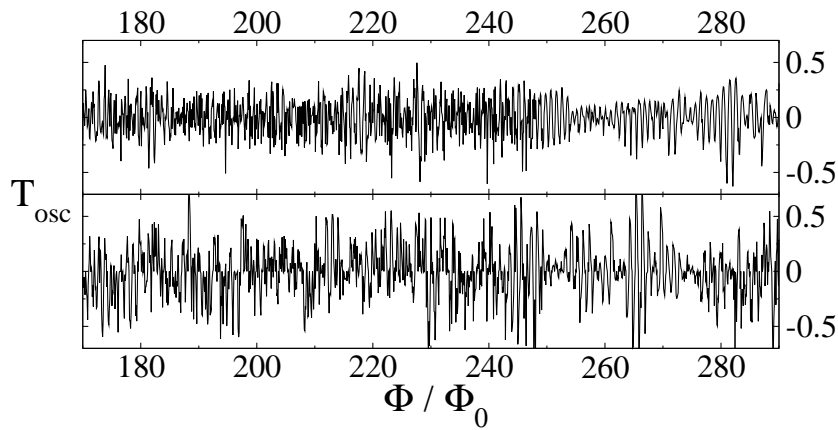}
\caption{Oscillating part of the magnetotransmission. The average transmission of one period around the corresponding value of $\Phi$ has been subtracted from $T(\Phi)$ 
to obtain $T_{\mathrm{osc}}(\Phi)$. Top panel: System of figure \ref{fig:ABlarge185}. Lower panel: Same system as in the top panel, but the outer radius of one ring arm is reduced from $63.7$\,nm to $61.9$\,nm which results in a difference in the arm widths of 10 percent.}
\label{fig:ABlarge}
\end{figure}

In figure \ref{fig:ABlarge185}, we show the magnetoconductance of a
large ring for a wide range of the magnetic flux, up to the quantum Hall
regime.  Up to a flux of about $300\,\Phi_0$, we see AB
oscillations with amplitudes approximately in the range $\Delta T \approx 0.5-1$
[see also figure \ref{fig:ABlarge}\,(a)].  For fluxes larger than
$300\,\Phi_0$, the oscillation amplitude becomes significantly smaller
($\Delta T \lesssim 0.5$), while for even higher values of $\Phi$ the
oscillations vanish completely and the transmission is equal to the
maximum value given by the number of quantum Hall edge channels in the leads.
This behavior can be understood by the following picture. The
cyclotron diameter $d_{\mathrm{c}}$ in (bulk) graphene is given by
\begin{equation}
\label{eq:cycd}
 d_{\mathrm{c}} = \frac{2E_F}{v_FeB}\,,
\end{equation}
where $E_F$ is the Fermi energy and $v_F$ is the Fermi velocity. In
our case $E_F=0.3\,t$ and the width of the ring arms is
$W_{\mathrm{A}} \approx 18$\,nm.  This means, that at a flux of
$\Phi\approx 230\,\Phi_0$, the cyclotron diameter is equal to an arm's
width. When $\Phi$ becomes comparable to this value, it becomes
more probable that the electron enters one arm than the other,
because the magnetic field dictates the direction of classical
propagation. Therefore, interference is reduced. 
When the magnetic field gets larger and larger, 
it becomes more and more unlikely for the electron to enter the second arm, 
and finally the interference vanishes when the
quantum Hall edge channels fully form. Since backscattering is
also strongly reduced at large magnetic fields, the transmission is only
limited by the number of propagating lead channels. This number
however is reduced for a fixed Fermi energy as the magnetic field is
increased (the distance between the Landau levels increases with $B$),
which leads to the step-like transmission in this regime. 
Furthermore, classical cyclotron effects are also important at the 
entrance to the ring. At
low fields, a large part of the wavefunction entering the ring is
simply reflected back into the lead by the inner wall of the ring.
As the field gets higher, the probability to turn around the corner into one of
the arms increases.  

\begin{figure}
\centering
\includegraphics[clip, width=11.0cm]{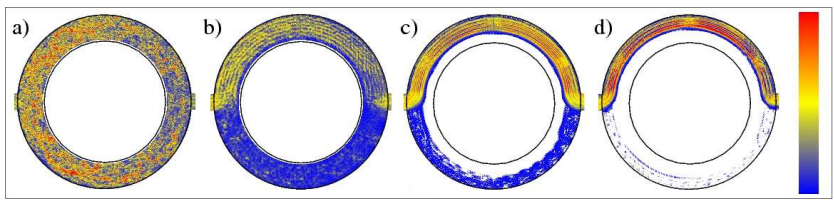}
\caption{Absolute value of the current density in the ring of figure \ref{fig:ABlarge185}. The electrons enter the ring from the left lead. Blue means low and red
means high current density.
(a) $\Phi=113\,\Phi_0$ (b) $\Phi=204\,\Phi_0$ (c) $\Phi=328\,\Phi_0$ (d) $\Phi=453\,\Phi_0$.}
\label{fig:current}
\end{figure}

This scenario is shown in figure \ref{fig:current}: the absolute value of the
current density in the ring is given for values of the flux $\Phi$ 
corresponding to the different regimes.
For $\Phi=113\,\Phi_0$ [figure \ref{fig:current}\,(a)] -- the
low transmission regime -- the current flows more or less equally
through both arms. As the flux is increased, the electrons are
forced more and more into the upper arm, and at very high fluxes,
when electrons are essentially travelling only along the upper arm,
interference gets suppressed and finally vanishes. Note that also the quantum
Hall edge channels become clearly visible at high fluxes. It is worth noting
that the small AB oscillations in the regime where $\Phi \gtrsim
300\,\Phi_0$ are not due to electrons entering the lower arm of the ring 
from the left lead. If this were true, the current in
figure \ref{fig:current}\,(c) would be localized at the inner border of
the lower arm, which is obviously not the case. In this regime, the
interference is due to paths going along the upper arm and leaving the
ring after a half circle interfering with paths which also start along 
the upper arm but which go around
the ring one more time before exiting. Such paths also give rise
to oscillations with period $\Phi_0$. Note that this effect is present only because the lead width in the chosen geometry is somewhat smaller than
the width of the ring, as then the quantum Hall edge channels in the ring arms may be scattered
at the lead opening and enter the opposite arm. Hence, this regime corresponds
to interferometry with quantum Hall edge channels. When the lead width 
is larger than the
ring arm width, scattering at the lead openings is suppressed, and no
interference is observed. The power spectra shown as insets in figure \ref{fig:ABlarge185} also 
support this picture: For low fields, several higher harmonics are present,
whereas at high fields only the fundamental frequency is visible because multiple
windings around the ring
are strongly suppressed in the quantum Hall regime.

The magnitude of the conductance seen in the quantum Hall regime -- an odd integer times $2e^2/h$ (see figure \ref{fig:ABlarge185}) -- is consistent with the bulk quantum Hall effect \cite{Novoselov2005, Zhang2005} as well as with theoretical results obtained for graphene nanoribbons \cite{Brey2006}. (Recall that in a two terminal geometry the conductance shows the quantization of $\sigma_{xy}$.) The steps of magnitude 2 are easily understood in terms of the valley degeneracy of the subbands in the leads. The quantization at odd rather than even integers is explained by a Berry phase effect involving the pseudospin associated with the sub-lattice degree of freedom \cite{CastroNeto2009}. 

In the experimental work of Ref.\,\cite{Russo2008}, a significant increase 
of the AB oscillation amplitude was observed at high fields when 
$W_{\mathrm{A}} \gtrsim d_{\mathrm{c}}$.
Note that our data in figure \ref{fig:ABlarge185} does not show any such 
increase (if anything, it shows the opposite). 
To investigate this further,
figure \ref{fig:ABlarge} shows the oscillating part of the
magnetoconductance for a range of $\Phi$ near the value where the
cyclotron diameter becomes equal to the width of a ring arm [around
$230\,\Phi_0$ according to (\ref{eq:cycd})]. The authors of
\cite{Russo2008} speculate that an asymmetry between the two ring arms could be
responsible for the increase of the oscillation amplitude. Our
numerical calculations, however, do not confirm this: for
figure \ref{fig:ABlarge}\,(b) we reduced the width of one arm by 10
percent, but also in this case, an inrease of the oscillation amplitude 
at high magnetic fields is absent.

The formation of quantum Hall edge channels
in a clean system happens for $d_\mathrm{c} \approx W_\mathrm{A}$, 
as observed in our simulations (figure \ref{fig:ABlarge185}). 
In the data of Ref.~\cite{Russo2008}, however, AB oscillations
are still visible even for much higher magnetic fields, when $d_\mathrm{c}<W_\mathrm{A}$.
Furthermore, despite being ten times larger than the rings considered in 
our numerical studies, the average conductance in the highly doped regime
is only on the order of a few $\frac{e^2}{h}$. These two features are a strong indication that these 
experiments are in the strongly disordered regime.

\begin{figure}
\centering
\includegraphics[clip, width=13.0cm]{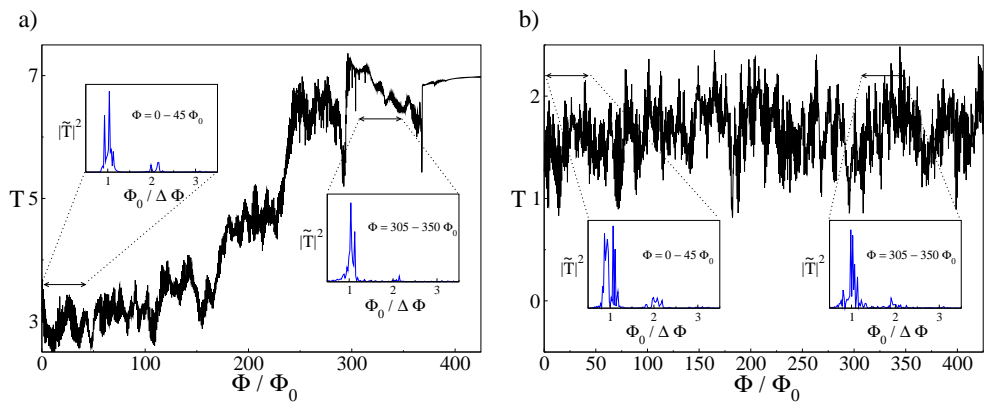}
\caption{Magnetotransmission of disordered rings with $R_1=45.5\,nm$,
$R_2=63.7\,nm$ and $W_L=13.5\,nm$
in the multi-mode regime (solid black lines). The insets show
corresponding power spectra. (a) Ring with edge disorder.
Here, we employed the etching disorder model of Ref.~\cite{Mucciolo2009}
 with $N_\mathrm{sweeps}=30$ and $p_i=0.05$ for $i=1\dots 30$ (for the
details on the parameters of this model, see Ref.~\cite{Mucciolo2009}).
(b) Ring with bulk disorder. For the bulk disorder, we choose
 a Gaussian disorder potential with $N_\mathrm{imp}/N_\mathrm{tot}=0.05$,
$\delta=0.1\,t$ and $\xi=2a$ (for the details
on the parameters of this model, see Ref.~\cite{Rycerz2007}).
Note that although the Gaussian disorder potential is smooth
on the lattice scale and hence does not scatter the valleys, valley
scattering nevertheless happens at the abrupt termination of the
lattice at the edges.}
\label{fig:disorder}
\end{figure}

Hence, we now turn to briefly discussing the effects of disorder on the AB oscillations. 
For this, we consider both edge disorder (employing the
model of edge disorder from Ref.~\cite{Mucciolo2009}) and bulk disorder
(using Gaussian disorder as in Ref.~\cite{Rycerz2007a}). Figure \ref{fig:disorder}(a) and (b)
show the magnetoconductance for edge and bulk disorder, respectively. In the case of 
edge disorder only, we still observe an increase of the transmission with magnetic field
due to classical cyclotron effects as in the clean case. In addition, the onset of
quantized Hall conductance is only shifted slightly to higher fields. In fact, since the
interior of the ring arms is ballistic, we still expect quantum Hall edge channels
to form for $d_\mathrm{c} \approx W_\mathrm{A}$. Moreover, as in the clean case
we observe interference of quantum Hall edge states for $\Phi \gtrsim 300\,\Phi_0 $. In contrast, for bulk disorder the 
transmission probability remains -- apart from conductance fluctuations -- virtually
unchanged as the magnetic field increases. In this case, the disorder broadens
the Landau levels enough to prevent the formation of edge channels.  
The similariy of this behavior to that of the experiments reported in Ref.~\cite{Russo2008} suggests that the experiments
are in the ``dirty limit'', dominated by bulk disorder. Such a conclusion is also
in agreement with Ref.~\cite{Molitor2009} where their experimental results 
for AB rings are explained in terms of diffusive metallic transport.

\begin{figure}
\centering
\includegraphics[clip, width=11.0cm]{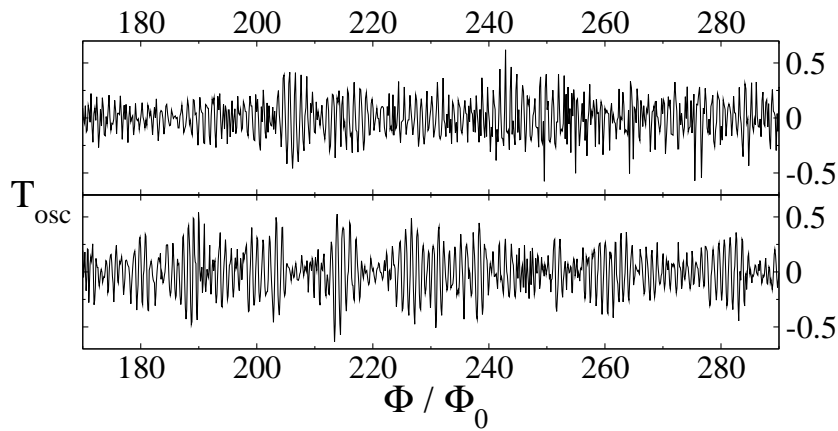}
\caption{Oscillating part of the magnetotransmission for disordered rings. Top panel: System
with edge disorder. Lower panel: System with Gaussian bulk disorder.
Disorder parameters are as in figure \ref{fig:disorder}.}
\label{fig:disorder2}
\end{figure}

Further characteristics of the diffusive limit include that the system 
remains isotropic for higher fields,
both ring arms carry an equal amount of current, and higher harmonics
are suppressed in comparison to the clean case (see figure \ref{fig:ABlarge185}).
Finally, figure \ref{fig:disorder2} shows the oscillating part of the magnetoconductance
in the vincinity of $d_\mathrm{c}\approx W_\mathrm{A}$. 
Again, the size of the oscillations
remains essentially unchanged with increasing magnetic flux.

A closer inspection of the oscillations reveals that in the case of edge
disorder, there are also aperiodic oscillations on top of the periodic
AB oscillations for $\Phi\approx 250-300\,\Phi_0$, where we observe the onset of edge
channel formation. These aperiodic oscillations are presumably due to
localization and resonant tunneling at the disordered edge, since there
is still backscattering among the not fully formed edge states. This
interpretation is supported by the fact that the aperiodic oscillations
are absent in the case of bulk disorder, where the edge channel
formation is completely suppressed. Furthermore, note that aperiodic
oscillations are also present (although to a less degree) for
$\Phi\approx 190-250\,\Phi_0$ in the clean rings, as the abrupt lattice termination also
serves as a source of \emph{local} disorder.

\section{Breaking the valley degeneracy in graphene rings}
\label{sec:Vdeg}

In the following, we demonstrate numerically that the lifting 
of the valley degeneracy of eigenstates in rings by 
a magnetic field perpendicular to the ring is also visible in 
transport properties of the ring. In Ref.~\cite{Recher2007}, 
the authors calculate the energy spectrum of a
closed graphene ring with {infinite mass boundary condition}
\cite{Berry1987} within the effective Dirac theory. The energy levels
$E_{\tau \pm}^{nm}$ are labeled by a radial quantum number $n$, an
angular momentum quantum number $m$, valley index $\tau$ and the
sign of the energy. At zero magnetic field, the authors find from their calculation the
degeneracy $E_{-\tau \pm}^{n(-m)} = E_{\tau \pm}^{nm}$, which is not
present when a perpendicular magnetic field is applied. It is worth to point out that this is due to
the fact that in addition to the usual time reversal (TR) operator
$\mathcal{T} = (\sigma_y\otimes\tau_y)\mathcal{C}$, there is another
antiunitary operator commuting with the Hamiltonian at zero magnetic
field, namely $\mathcal{T}_{\mathrm{v}} =
(\sigma_y\otimes\tau_x)\mathcal{C}$. In contrast to $\mathcal{T}$,
$\mathcal{T}_{\mathrm{v}}$ is associated with a symplectic symmetry,
since $\mathcal{T}_{\mathrm{v}}^2=-1$. This ensures the degeneracy of
both valleys (Kramer's degeneracy \cite{Messiah1970}). Note that this reasoning holds only if the boundary condition preserves the
symmetry connected with $\mathcal{T}_{\mathrm{v}}$; such is the case
for the infinite mass boundary condition used in Ref.~\cite{Recher2007} but is not the case, for
example, for valley mixing armchair boundaries. Once a magnetic field
pierces the ring, the $\mathcal{T}_{\mathrm{v}}$ symmetry is broken 
and the valley degeneracy is
lifted. This mechanism has been numerically observed in ballistic
graphene quantum dots with smooth mass confinement \cite{Wurm2009},
where the amplitude of universal conductance fluctuations was
considered.  Here we follow a different idea to probe the effect of
breaking valley degeneracy with a magnetic field,
namely we calculate numerically the conductance of a ring weakly
coupled to two leads. The conductance is then expected to be peaked at
values of the Fermi energy that match the energies of eigenstates
of the closed ring.

\begin{figure}
\centering
\includegraphics[clip, width=9.0cm]{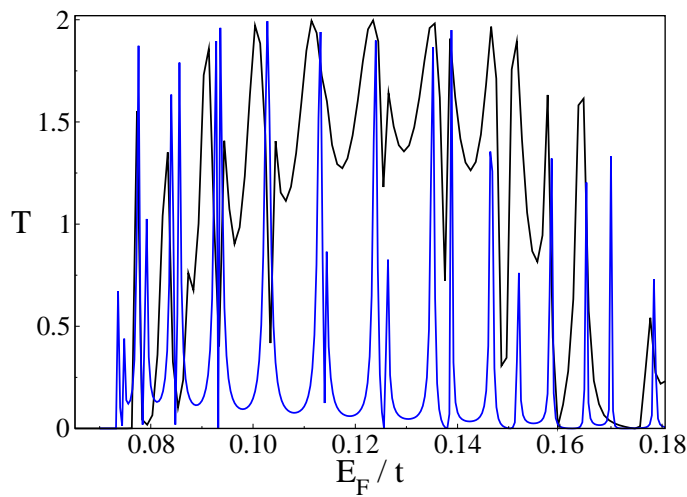}
\caption{Transmission of rings with smooth mass term ($\omega = 0.05\,\sqrt{t}/a$) and $R_1=9.84$\,nm, $R_2=19.7$\,nm, $W_L=9.84$\,nm, with leads
that are strongly (black) and weakly (blue) coupled to the system as a function of the Fermi energy in the two-mode regime (transverse lead channels are twofold 
valley degenerate).}
\label{fig:Couplings}
\end{figure}

In order to have valley degenerate states at zero magnetic field, we
have to avoid scattering at local armchair boundaries, as mentioned
above.  To model the infinite mass boundary numerically, we use
smooth mass confinement as introduced in \cite{Wurm2009}. Due to the
finite escape time of electrons injected into the ring from one lead,
the effects of inter-valley scattering will be suppressed in such a
system, provided the mass confinement is smooth enough. We use a mass
term that is zero in the middle of the ring and increases
quadratically towards the boundaries [cf.
figure \ref{fig:ringScheme2}\,(a)]: $m(x,y) = \omega^2\,\delta^2(x,y)/2$,
where $\delta(x,y) = \sqrt{x^2+y^2}-\bar{R}$ for a ring with average radius $\bar{R}$ centered at $x=y=0$. 
In the leads we use a
consistent mass term, with $\delta(x,y) = y$ for leads in the $x$-direction. 
In the region where the leads are attached, we match both terms in
a smooth way. In this region we also introduce a mass-induced barrier, 
that can be varied in height, in order to 
tune the coupling strength between leads and ring. 

Figure \ref{fig:Couplings} shows the transmission of rings with both strong
and and weak coupling as a function of the Fermi energy
$E_F$. We checked that the transmission is insensitive to the
orientation of the system with respect to the lattice orientation at
the energies considered. This means that the detailed edge structure
of ring and leads do not play a role, and implies that the 
mass confinement is smooth enough to suppress inter-valley scattering. 
Upon reducing the coupling strength,
the oscillations in the transmission of the ring with strongly coupled
leads turn into sharp peaks. The system studied is a ring with 
$R_1=9.84$\,nm, $R_2=19.7$\,nm and width
$W_L=9.84$\,nm. Note that the effective width of leads and ring is
reduced due to the mass confinement. The mass potential is $0.5\,t$ at the physical borders 
of the system for the value of $\omega$ used (see caption of figure \ref{fig:Couplings}).

\begin{figure}
\centering
\includegraphics[clip, width=10.0cm]{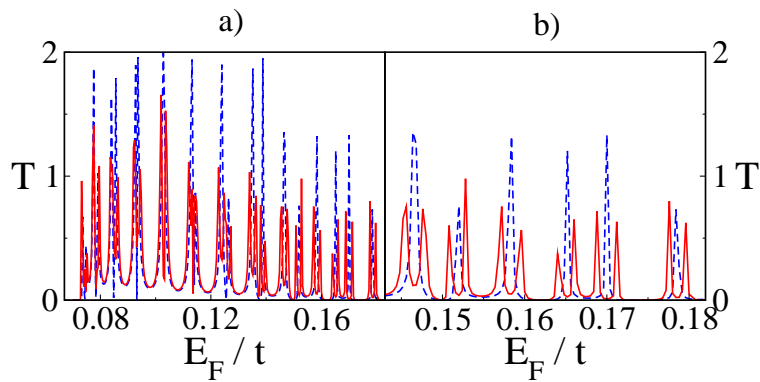}
\caption{Transmission of the weakly coupled ring from figure \ref{fig:Couplings} for zero flux (blue dashed lines) and a flux $\Phi=0.17\,\Phi_0$ (red solid lines).
(b) shows a zoom on the 6 (or 12, respectively) rightmost peaks of (a).}
\label{fig:BrokenDeg_2}
\end{figure}
To find out, wether the energy spectrum of our ring shows the valley
degeneracy connected with the symmetry described by
$\mathcal{T}_{\mathrm{v}}$, we try to break this supposed degeneracy
by applying a magnetic field perpendicular to the ring.
Figure \ref{fig:BrokenDeg_2}\,(a) shows the numerical results, where
the magnetic flux through the area
$\pi \bar{R}^2$ with $\bar{R}=14.8$\,nm is $\Phi=0$ (dashed lines) and
$\Phi=0.17\,\Phi_0$ (solid lines), respectively. We find that
the transmission peaks split and that their
heights are in general reduced. This is the behavior expected
from degeneracy lifting, as the weight of each split peak 
now corresponds to only a single, non-degenerate level. Note that 
for $E<0.14\,t$ the peaks partially overlap; the splitting of the 
valley degeneracy is particularly clear for $E>0.14\,t$ where the peaks are well seperated, 
as shown in figure \ref{fig:BrokenDeg_2}\,(b). These results show
unambiguously that the lifting of the valley-degeneracy by a magnetic flux
in weakly coupled graphene rings can also be observed in the transport properties.

\section{Conclusion}

In this work we have investigated two different aspects of mesoscopic graphene rings:
First, we have numerically studied the Aharonov-Bohm effect in rings cut out 
of the graphene lattice. Both small rings with a single mode as well as 
large rings supporting many modes exhibit clear $h/e$ Aharonov-Bohm oscillations. 
Although disorder has a strong influence on the average transmission
of graphene rings especially in the high field regime, the Aharonov-Bohm 
oscillations are influenced only very little by both edge and bulk disorder.
Second, we have shown that the signature of the splitting of the valley-degeneracy
by a magnetic field in rings with a mass confinement can also be observed
in the transport properties of rings that are weakly coupled to leads.

\ack JW acknowledges support from Deutsche Forschungsgemeinschaft within GRK 638, and
MW and KR support from Deutsche Forschungsgemeinschaft within SFB 689.
The work at Duke was supported in part by the U.S. NSF (DMR-0506953) and the DAAD.
We would like to thank \.Inan\c{c} Adagideli, Patrik Recher, Adam Rycerz and Bj\"orn Trauzettel
for helpful discussions.\\


\begin{thebibliography}{10}

\bibitem{Geim2007}
A.~K. Geim and K.~S. Novoselov.
\newblock The rise of graphene.
\newblock {\em Nature Mat.}, 6(3):183--191, March 2007.

\bibitem{CastroNeto2009}
A.~H. Castro~Neto, F.~Guinea, N.~M.~R. Peres, K.~S. Novoselov, and A.~K. Geim.
\newblock The electronic properties of graphene.
\newblock {\em Rev. Mod. Phys.}, 81(1):109, 2009.

\bibitem{Tworzydlo2006}
J.~Tworzyd\l{}o, B.~Trauzettel, M.~Titov, A.~Rycerz, and C.~W.~J. Beenakker.
\newblock Sub-{P}oissonian shot noise in graphene.
\newblock {\em Phys. Rev. Lett.}, 96(24):246802, 2006.

\bibitem{Danneau2008}
R.~Danneau, F.~Wu, M.~F. Craciun, S.~Russo, M.~Y. Tomi, J.~Salmilehto, A.~F.
  Morpurgo, and P.~J. Hakonen.
\newblock Shot noise in ballistic graphene.
\newblock {\em Phys. Rev. Lett.}, 100(19):196802, 2008.

\bibitem{DiCarlo2008}
L.~DiCarlo, J.~R. Williams, Yiming Zhang, D.~T. McClure, and C.~M. Marcus.
\newblock Shot noise in graphene.
\newblock {\em Phys. Rev. Lett.}, 100(15):156801, 2008.

\bibitem{Beenakker2008}
C.~W.~J. Beenakker.
\newblock Colloquium: {A}ndreev reflection and {K}lein tunneling in graphene.
\newblock {\em Rev. Mod. Phys.}, 80(4):1337, 2008.

\bibitem{Suzuura2002}
H.~Suzuura and T.~Ando.
\newblock Crossover from symplectic to orthogonal class in a two-dimensional
  honeycomb lattice.
\newblock {\em Phys. Rev. Lett.}, 89(26):266603, Dec 2002.

\bibitem{Khveshchenko2006}
D.~V. Khveshchenko.
\newblock Electron localization properties in graphene.
\newblock {\em Phys. Rev. Lett.}, 97(3):036802, 2006.

\bibitem{McCann2006}
E.~McCann, K.~Kechedzhi, Vladimir~I. Fal'ko, H.~Suzuura, T.~Ando, and B.~L.
  Altshuler.
\newblock Weak-localization magnetoresistance and valley symmetry in graphene.
\newblock {\em Phys. Rev. Lett.}, 97(14):146805, 2006.

\bibitem{Morpurgo2006}
A.~F. Morpurgo and F.~Guinea.
\newblock Intervalley scattering, long-range disorder, and effective
  time-reversal symmetry breaking in graphene.
\newblock {\em Phys. Rev. Lett.}, 97(19):196804, 2006.

\bibitem{Aleiner2006}
I.~L. Aleiner and K.~B. Efetov.
\newblock Effect of disorder on transport in graphene.
\newblock {\em Phys. Rev. Lett.}, 97(23):236801, 2006.

\bibitem{Ponomarenko2008}
L.~A. Ponomarenko, F.~Schedin, M.~I. Katsnelson, R.~Yang, E.~W. Hill, K.~S.
  Novoselov, and A.~K. Geim.
\newblock Chaotic {D}irac billiard in graphene quantum dots.
\newblock {\em Science}, 320(5874):356--358, 2008.

\bibitem{Wurm2009}
J.~Wurm, A.~Rycerz, {I}. Adagideli, M.~Wimmer, K.~Richter, and H.~U. Baranger.
\newblock Symmetry classes in graphene quantum dots: Universal spectral
  statistics, weak localization, and conductance fluctuations.
\newblock {\em Phys. Rev. Lett.}, 102(5):056806, 2009.

\bibitem{Aharonov1959}
Y.~Aharonov and D.~Bohm.
\newblock Significance of electromagnetic potentials in the quantum theory.
\newblock {\em Phys. Rev.}, 115(3):485--491, Aug 1959.

\bibitem{Webb1985}
R.~A. Webb, S.~Washburn, C.~P. Umbach, and R.~B. Laibowitz.
\newblock Observation of $h/e$ {A}haronov-{B}ohm oscillations in normal-metal
  rings.
\newblock {\em Phys. Rev. Lett.}, 54(25):2696--2699, Jun 1985.

\bibitem{Washburn1986}
S.~Washburn and R.~A. Webb.
\newblock {A}haronov-{B}ohm effect in normal metal quantum coherence and
  transport.
\newblock {\em Adv. Phys.}, 35(4):375--422, 1986.

\bibitem{Russo2008}
S.~Russo, J.~B. Oostinga, D.~Wehenkel, H.~B. Heersche, S.~S. Sobhani, L.~M.~K.
  Vandersypen, and A.~F. Morpurgo.
\newblock Observation of {A}haronov-{B}ohm conductance oscillations in a
  graphene ring.
\newblock {\em Phys. Rev. B}, 77(8):085413, 2008.

\bibitem{Molitor2009}
F.~Molitor, M.~Huefner, A.~Jacobsen, A.~Pioda, C.~Stampfer, K.~Ensslin, and
  T.~Ihn.
\newblock {A}haronov-{B}ohm effect in a side-gated graphene ring.
\newblock arXiv:0904.1364v1, 2009.

\bibitem{Rycerz2009}
A.~Rycerz.
\newblock {A}haronov-{B}ohm effect and valley polarization in nanoscopic
  graphene rings.
\newblock {\em Acta Phys. Pol. A}, 115(1):322--325, 2009.

\bibitem{Berry1987}
M.~V. Berry and R.~J. Mondragon.
\newblock Neutrino billards: {T}ime-reversal symmetry-breaking without magnetic
  fields.
\newblock {\em Proc. R. Soc. Lond. A}, 412:53--74, 1987.

\bibitem{Recher2007}
P.~Recher, B.~Trauzettel, A.~Rycerz, Ya.~M. Blanter, C.~W.~J. Beenakker, and
  A.~F. Morpurgo.
\newblock {A}haronov-{B}ohm effect and broken valley degeneracy in graphene
  rings.
\newblock {\em Phys. Rev. B}, 76(23):235404, 2007.

\bibitem{Wimmer2008a}
M.~Wimmer and K.~Richter.
\newblock Optimal block-tridiagonalization of matrices for coherent charge
  transport.
\newblock arXiv:0806.2739v1, 2008.

\bibitem{Novoselov2005}
K.~S. Novoselov, A.~K. Geim, S.~V. Morozov, D.~Jiang, M.~I. Katsnelson, I.~V.
  Grigorieva, S.~V. Dubonos, and A.~A. Firsov.
\newblock Two-dimensional gas of massless {D}irac fermions in graphene.
\newblock {\em Nature}, 438(7065):197--200, November 2005.

\bibitem{Zhang2005}
Y.~Zhang, Y.-W. Tan, H.~L. Stormer, and P.~Kim.
\newblock Experimental observation of the quantum {H}all effect and {B}erry's
  phase in graphene.
\newblock {\em Nature}, 438(7065):201--204, 2005.

\bibitem{Brey2006}
L.~Brey and H.~A. Fertig.
\newblock Edge states and the quantized {H}all effect in graphene.
\newblock {\em Phys. Rev. B}, 73(19):195408, 2006.

\bibitem{Mucciolo2009}
E.~R. Mucciolo, A.~H. Castro~Neto, and C.~H. Lewenkopf.
\newblock Conductance quantization and transport gaps in disordered graphene
  nanoribbons.
\newblock {\em Phys. Rev. B}, 79(7):075407, 2009.

\bibitem{Rycerz2007}
A.~Rycerz, J.~Tworzydlo, and C.~W.~J. Beenakker.
\newblock Valley filter and valley valve in graphene.
\newblock {\em Nature Phys.}, 3(3):172--175, March 2007.

\bibitem{Rycerz2007a}
A.~Rycerz, J.~Tworzydlo, and C.~W.~J. Beenakker.
\newblock Anomalously large conductance fluctuations in weakly disordered
  graphene.
\newblock {\em Europhys. Lett.}, 79(5):57003, 2007.

\bibitem{Messiah1970}
A.~Messiah.
\newblock {\em Quantum Mechanics 2}.
\newblock North-Holland, 1970.

\end{thebibliography}
\end{document}